\documentclass[prl,twocolumn,superscriptaddress]{revtex4-1}
\usepackage{}
\usepackage{amssymb}
\usepackage{amsfonts}
\usepackage{bbm}
\usepackage{mathrsfs}
\usepackage{graphics,graphicx,epsfig,bm,amsmath,amsthm,amssymb}
\usepackage{bm}
\usepackage{bbm}
\usepackage{longtable}
\usepackage{multirow}
\usepackage{array}
\usepackage{color}
\usepackage{diagbox}
\usepackage{amsmath}
\usepackage{upgreek}
\usepackage{multirow}
\usepackage{threeparttable}
\usepackage{braket}
\usepackage[usenames,dvipsnames]{xcolor}

\usepackage{float}
\usepackage[a4paper,colorlinks=true,
linkcolor=blue,citecolor=blue,
pdfauthor={ },
pdftitle={ },
pdfsubject={ },
pdfkeywords={ }]{hyperref}

\begin{document}
	
\title{Experimental investigation of coherent ergotropy in a single spin system}
\affiliation{CAS Key Laboratory of Microscale Magnetic Resonance and School of Physical Sciences, University of Science and Technology of China, Hefei 230026, China}
\affiliation{CAS Center for Excellence in Quantum Information and Quantum Physics, University of Science and Technology of China, Hefei 230026, China}
\affiliation{Hefei National Laboratory, University of Science and Technology of China, Hefei 230088, China}
\affiliation{Anhui Province Key Laboratory of Scientific Instrument Development and Application, University of Science and Technology of China, Hefei 230026, China}
\affiliation{Institute of Quantum Sensing and School of Physics, Zhejiang University, Hangzhou 310027, China}

\author{Zhibo Niu}
\affiliation{CAS Key Laboratory of Microscale Magnetic Resonance and School of Physical Sciences, University of Science and Technology of China, Hefei 230026, China}
\affiliation{CAS Center for Excellence in Quantum Information and Quantum Physics, University of Science and Technology of China, Hefei 230026, China}
\affiliation{Anhui Province Key Laboratory of Scientific Instrument Development and Application, University of Science and Technology of China, Hefei 230026, China}

\author{Yang Wu}
\affiliation{CAS Key Laboratory of Microscale Magnetic Resonance and School of Physical Sciences, University of Science and Technology of China, Hefei 230026, China}
\affiliation{CAS Center for Excellence in Quantum Information and Quantum Physics, University of Science and Technology of China, Hefei 230026, China}
\affiliation{Anhui Province Key Laboratory of Scientific Instrument Development and Application, University of Science and Technology of China, Hefei 230026, China}

\author{Yunhan Wang}
\affiliation{CAS Key Laboratory of Microscale Magnetic Resonance and School of Physical Sciences, University of Science and Technology of China, Hefei 230026, China}
\affiliation{CAS Center for Excellence in Quantum Information and Quantum Physics, University of Science and Technology of China, Hefei 230026, China}
\affiliation{Hefei National Laboratory, University of Science and Technology of China, Hefei 230088, China}
\affiliation{Anhui Province Key Laboratory of Scientific Instrument Development and Application, University of Science and Technology of China, Hefei 230026, China}

\author{Xing Rong}
\email{xrong@ustc.edu.cn}
\affiliation{CAS Key Laboratory of Microscale Magnetic Resonance and School of Physical Sciences, University of Science and Technology of China, Hefei 230026, China}
\affiliation{CAS Center for Excellence in Quantum Information and Quantum Physics, University of Science and Technology of China, Hefei 230026, China}
\affiliation{Hefei National Laboratory, University of Science and Technology of China, Hefei 230088, China}
\affiliation{Anhui Province Key Laboratory of Scientific Instrument Development and Application, University of Science and Technology of China, Hefei 230026, China}

\author{Jiangfeng Du}
\affiliation{CAS Key Laboratory of Microscale Magnetic Resonance and School of Physical Sciences, University of Science and Technology of China, Hefei 230026, China}
\affiliation{CAS Center for Excellence in Quantum Information and Quantum Physics, University of Science and Technology of China, Hefei 230026, China}
\affiliation{Hefei National Laboratory, University of Science and Technology of China, Hefei 230088, China}
\affiliation{Anhui Province Key Laboratory of Scientific Instrument Development and Application, University of Science and Technology of China, Hefei 230026, China}
\affiliation{Institute of Quantum Sensing and School of Physics, Zhejiang University, Hangzhou 310027, China}

	\begin{abstract}
		
		 Ergotropy is defined as the maximum amount of work that can be extracted through a unitary cyclic evolution. It plays a crucial role in assessing the work capacity of a quantum system. Recently, the significance of quantum coherence in work extraction has been theoretically identified, revealing that quantum states with more coherence possess more ergotropy compared to their dephased counterparts. 
However, an experimental study of the coherent ergotropy remains absent. 
Here, we report an experimental investigation of the coherent ergotropy in a single spin system.
Based on the method of measuring ergotropy with an ancilla qubit, both the coherent and incoherent components of the ergotropy for the non-equilibrium state were successfully extracted. 
The increase in ergotropy induced by the increase in the coherence of the system was observed by varying the coherence of the state. 
Our work reveals the interplay between quantum thermodynamics and quantum information theory, future investigations could further explore the role other quantum attributes play in thermodynamic protocols.
		
	\end{abstract}

	\maketitle

Quantum thermodynamics bridges two pillars of physics: quantum mechanics and thermodynamics\cite{Book1,Review1,Review2,NP1}. 
One of its central topics is the work extraction from an out-of-equilibrium system. As a fundamental process in the thermodynamics of quantum systems, the work extraction has been extensively studied\cite{Extract1,Extract2,Extract3,Extract4,Extract5,Extract6,Extract7,Extract8}.
The concept of ergotropy, defined as the maximal amount of work that is extractable via cyclic unitary evolution, was brought up to describe the work capacity of a quantum state\cite{Erg_Definition}. Beyond the mean energy of a quantum state, ergotropy reflects how much usable energy a quantum system can deliver to external systems. It has been measured recently in several experiments to showcase the performance of their thermodynamic devices\cite{Exp_NMR,Exp_SingleAtom,Exp_Flywheel}. 
The connection between the ergotropy of a quantum state and its quantum signatures has been identified theoretically\cite{Erg_Coh1,Erg_Coh2,Erg_Coh3,Erg_Coh4,Erg_Correlation1,Erg_Correlation2,Erg_Correlation3,Erg_Entanglement1,Erg_Entanglement2,Erg_Entanglement3,Erg_QuantumSignatures}. One of the most fundamental non-classical features of a quantum system is the coherence, its contribution to the ergotropy is isolated by dividing the optimal working-extracting operation into a coherence-preserving cycle and a coherence-consuming one\cite{Erg_Coh1,Erg_Coh2,Resource_Theo}. 
Despite the theoretical advancements, an experimental investigation to demonstrate how coherence yields larger ergotropy remains absent. 
It is essential to experimentally study the relationship between coherence and ergotropy, which provides insights to both theoretical investigation and potential applications in thermodynamic protocols. 

Here, we report an experimental investigation of the coherent ergotropy in a single spin system. 
We developed a method for ergotropy measurement with an ancilla qubit, which avoids the usage of complicated quantum state tomography. 
An isotopically purified diamond ([$^{13}$C] = 0.001\%) was synthesized so that the electron spin in nitrogen-vacancy (NV) center\cite{NV} with sufficiently long coherence time can be used to demonstrate the relationship between ergotropy and coherence (see the Supplementary Material\cite{Supple} for detailed information on the impact of decoherence on experimental results).
In our experiment, the coherent and incoherent components of the ergotropy were extracted separately by dividing the optimal operation that extracts the ergotropy of the state into a coherence-preserving operation and a coherence-consuming one. 
Moreover, by adding coherence into a totally dephased state in energy basis, we observed a corresponding increase of the coherent ergotropy. Thus, the contribution of the coherence to the ergotropy was systematically revealed.

We study the work extraction process by considering a quantum system in state $\rho$ subjected to Hamiltonian $H_S=\sum_{n} \epsilon_{n}|\epsilon_{n}\rangle\langle\epsilon_{n}|$, where $\epsilon_{n}$ ($|\epsilon_{n}\rangle$) is the energy eigenvalue (eigenstate). The mean energy can be evaluated simply as $\langle H_S\rangle_{\rho}=\rm{Tr}$$[\rho H_S]$. However, this is not the quantity of work that can be extracted to external systems. Thus, the ergotropy of a quantum state, which is defined as the maximal amount of extractable work under cyclic unitary evolution, is introduced to quantify the usable work of a quantum state. We consider a cyclic evolution in which the total Hamiltonian of the system can be written as $H_{total}=H_S+V_{ext}(t)\ (0\leq t \leq T)$, with $V_{ext}(t)$ being the external driving or coupling to other systems, it satisfies $V_{ext}(0)=V_{ext}(T)=0$. The work extracted is $W=\mathrm{Tr}[\mathit{\rho H_S}]-\mathrm{Tr}[\mathit{U\rho U^{\dagger} H_S}]$, where $U$ is the evolution operator corresponding to the work extraction protocol. Because any unitary evolution can be generated through a suitable choice of $V_{ext}(t)$\cite{Unitary_Operation}, the ergotropy can be found by taking the maximum of the extracted work over all possible evolution operators:
\begin{equation}
	\mathcal{E}(\rho)=\max_{U} (\mathrm{Tr}[\mathit{\rho H_S}]-\mathrm{Tr}[\mathit{U\rho U^{\dagger} H_S}]).
\end{equation}
The optimal operation is denoted as $E_{\rho}$ and the corresponding final state is $P_{\rho}=E_{\rho}\rho E_{\rho}^{\dagger}$. $P_{\rho}$ is called the passive state of $\rho$, indicating that no additional work can be extracted via any unitary operations\cite{Passive_State}. 

\begin{figure}\centering
	\includegraphics[width=1\columnwidth]{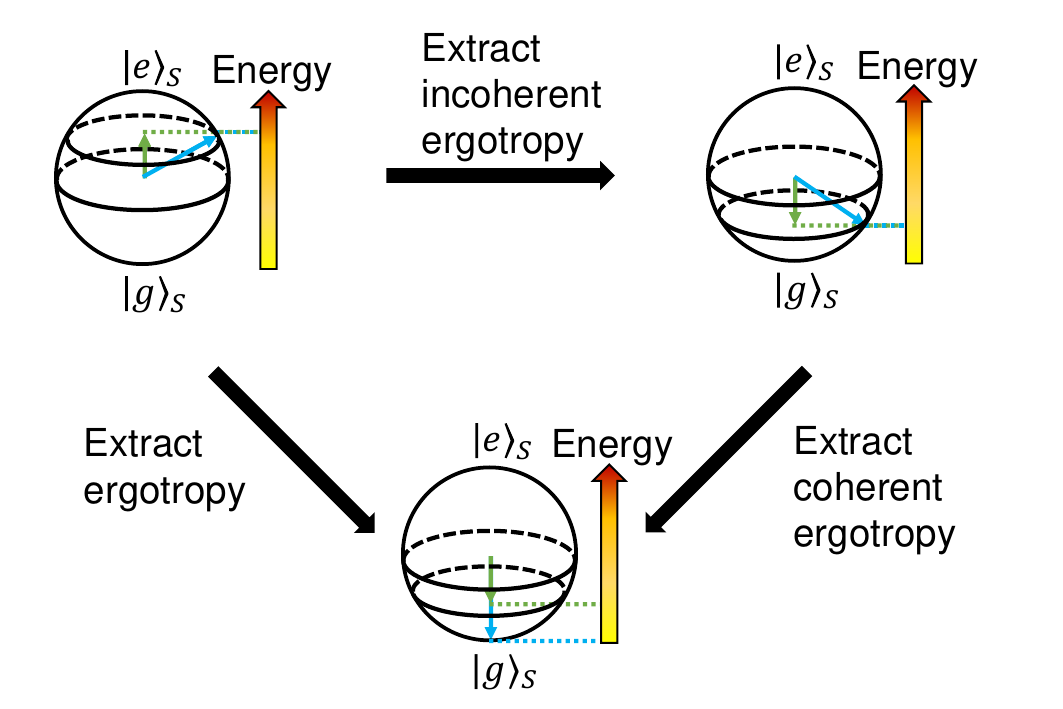}
	\caption{\label{fig1}
		Schematic diagram of the optimal extractions in a Bloch sphere in energy basis. $|e\rangle_S$ corresponds to the high-energy level. The energy of initial state with coherence (blue) and its totally dephased counterpart (green) can be extracted by first extracting the incoherent ergotropy and then the coherent part or directly extracting the ergotropy. The totally dephased state yields no contribution in the coherent extraction process.
	}
\end{figure}

To establish the connection between the coherence and the work extraction, the coherence of state $\rho$ is quantified by the quantum relative entropy of $\rho$ and its totally dephased state in energy basis\cite{Coh_Definition1,Coh_Definition2}: $C(\rho)=D(\rho||\delta_{\rho})=\mathrm{Tr}[\mathit{\rho(\log{\rho}-\log{\delta_{\rho}})}]$, where $\delta_{\rho}=\sum_{n}|\epsilon_{n}\rangle\langle\epsilon_{n}|\langle\epsilon_{n}|\rho|\epsilon_{n}\rangle$.
Subsequently, the incoherent component of the ergotropy is defined as the maximum work extractable from $\rho$ without altering its coherence. Specifically, the incoherent ergotropy is given by $\mathcal{E}_{i}(\rho)=\max_{V} (\mathrm{Tr}[\mathit{\rho H_S}]-\mathrm{Tr}[\mathit{V\rho V^{\dagger} H_S}])$, where $V$ is a unitary transformation that satisfies $C(\rho)=C(V\rho V^{\dagger})$\cite{Coh_Definition2,Incoh_Operation1,Incoh_Operation2}. The optimal coherence-preserving operation that extracts maximum work is denoted as $V_{\pi}$, where $\pi$ signifies the permutation of energy basis up to irrelevant phase factors. If the eigenvalues of Hamiltonian $H_S$ is in a descending order, $V_{\pi}$ rearranges the population of $\rho$ in an ascending order. Therefore, state $\rho$ and its totally dephased counterpart $\delta_{\rho}$ have the same amount of incoherent ergotropy, as they share the same population distribution in energy basis\cite{Equal_Incoh}. The resulting state after the optimal incoherent extraction is denoted as $\sigma_{\rho}=V_{\pi}\rho V_{\pi}^{\dagger}$. State $\sigma_{\rho}$ maintains the amount of coherence of $\rho$ but stores less energy. To extract the rest part of its remaining extractable energy, the operations that alter its coherence is introduced. The coherent ergotropy of $\rho$ is thus isolated as the maximum extractable work of $\sigma_{\rho}$: $\mathcal{E}_{c}(\rho) = \mathcal{E}(\sigma_{\rho})$. The process of extracting the ergotropy of $\rho$ is equivalent to a sequential extraction of its incoherent and coherent ergotropy, formalized as
\begin{equation}
	\mathcal{E}_{c}(\rho)=\mathcal{E}(\rho)-\mathcal{E}_{i}(\rho).
	\label{Coh_Erg}
\end{equation} 
The relation of the coherence and the coherent ergotropy can be quantitively expressed by introducing an inverse temperature parameter $\beta$ and a Gibbs state $\rho_{\beta}$: $\beta \mathcal{E}_c(\rho) = C(\rho)+D(P_{\delta_{\rho}}||\rho_{\beta})-D(P_{\rho}||\rho_{\beta})$\cite{Erg_Coh1}, where $P_{\delta_{\rho}}$ is the passive state of $\delta_{\rho}$, this relation is valid for every finite $\beta$.


We study the coherent ergotropy in energy basis such that the system Hamiltonian is diagonal.
Without loss of generality, we choose our model Hamiltonian as 
$H_S =\begin{pmatrix}
	\epsilon & 0\\
	0 & 0\\
\end{pmatrix}$,
where $\epsilon=1.05\ \mathrm{MHz}$. The initial system state is chosen as $\rho_{S}=|\psi_0\rangle\langle\psi_0|$, $|\psi_0\rangle=(\sqrt{2}|e\rangle_S+|g\rangle_S)/\sqrt{3}$, and its totally dephased state is $\delta_{\rho_S}=(2|e\rangle_S {}_S\langle e|+|g\rangle_S {}_S\langle g|)/3$. 
The states and the ergotropic extractions are depicted on Bloch spheres in Fig.~\ref{fig1}. The blue (green) arrows represent $\rho_{S}$ ($\delta_{\rho_S}$) and the states after it is manipulated.
Two approaches to the passive state are shown in Fig.~\ref{fig1}.
The first one is directly applying the optimal operation that transforms $\rho_{S}$ ($\delta_{{\rho_S}}$) into its passive state $P_{\rho_{S}}$ ($P_{\delta_{{\rho_S}}}$). The total ergotropy of the initial system state is extracted, and the coherence of $\rho_{S}$ is consumed completely at the same time. The total ergotropy of the two states, $\mathcal{E}(\rho_{S})$ and $\mathcal{E}(\delta_{{\rho_S}})$, are obtained.
The second approach to the passive states consists of two steps. 
The optimal incoherent operation $V_{\pi}=|e\rangle_S{}_S\langle g|-|g\rangle_S{}_S\langle e|$ is firstly applied to swap the order of population in energy basis. 
The resulting state after the incoherent extraction of $\rho_{S}$ still stores the coherence and the coherent ergotropy of $\rho_{S}$. Meanwhile, the totally dephased state possesses zero extractable work after the incoherent extraction. The second step is to apply a coherence-consuming operation that transforms $\sigma_{{\rho_S}}$ into $P_{\rho_{S}}$ to extract the coherent ergotropy of $\rho_S$. Meanwhile, the dephased state remains unchanged and no work is extracted from it.

\begin{figure}\centering
	\includegraphics[width=1\columnwidth]{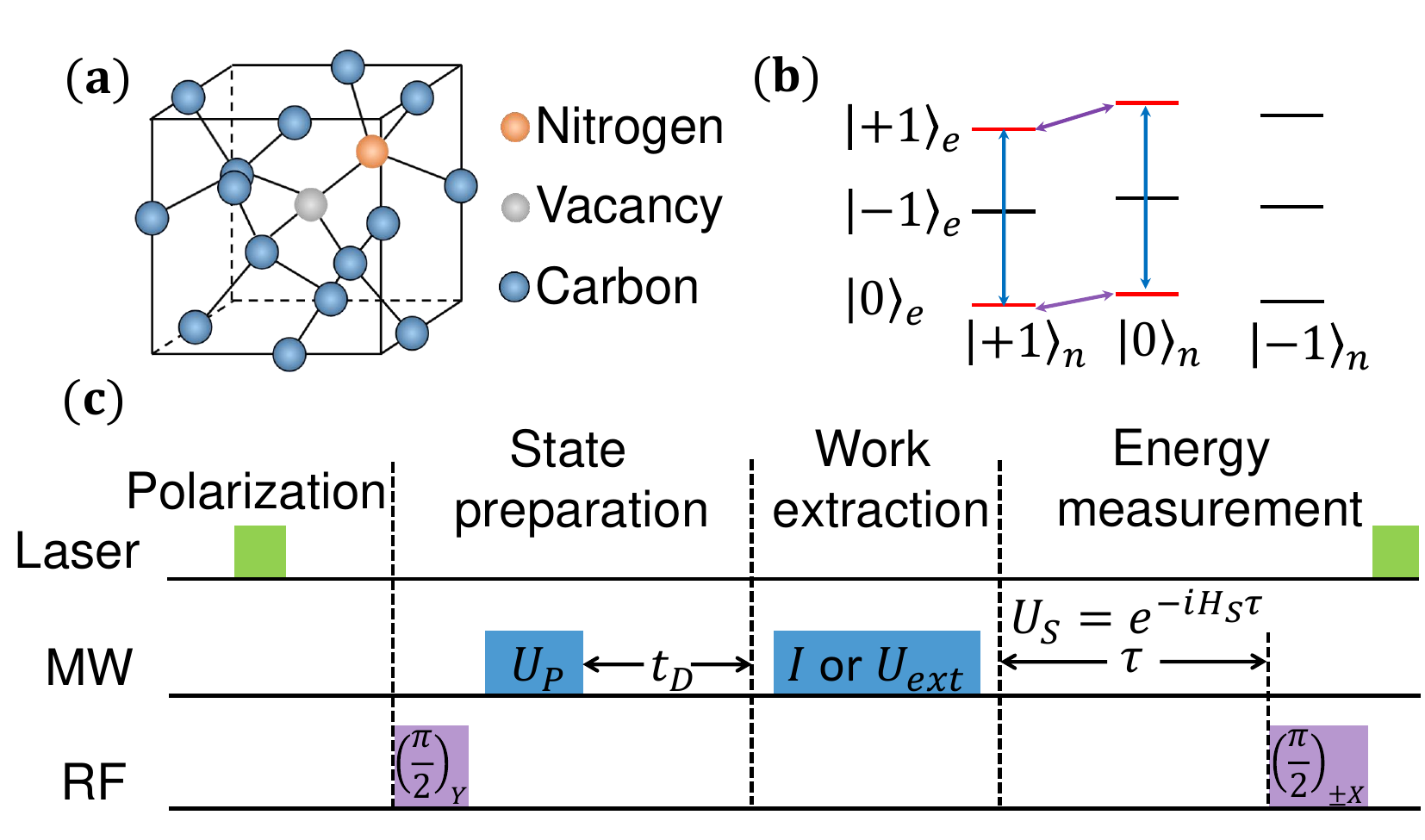}
	\caption{\label{fig2}
		The NV center system and the experimental pulse sequence.
		(a) Schematic atomic structure of the NV center.
		(b) Ground state energy levels of the NV center. The red lines denote the energy levels utilized in this experiment. The transitions between different electron (nuclear) spin states can be steered by microwave (radio-frequency) pulses represented by blue (purple) arrows.
		(c) Experimental pulse sequence of ergotropy measurement, which includes polarization, state preparation, work extraction and energy measurement. MW pulses are applied to prepare the electron state and extract work from it. RF pulses transform the nuclear spin state to assist the ergotropy measurement.
	}
\end{figure}

A single NV center in diamond was utilized to investigate the coherent ergotropy of a quantum state. The NV center consists of a substitutional nitrogen atom adjacent to a vacancy shown in Fig.~\ref{fig2}(a). When an external magnetic field is applied along the symmetric axis of the NV center, the Hamiltonian of the NV center can be written as
\begin{equation}
	H_{NV} = 2\pi(DS_{z}^{2}+\omega_{e}S_{z}+QI_{z}^{2}+\omega_{n}I_{z}+AS_{z}I_{z}),
\end{equation}
where $S_{z}$ ($I_{z}$) is the spin-1 operator of the electron (nuclear) spin, $D=2.87\ \mathrm{GHz}$ is the electronic zero-field splitting, $Q=-4.95\ \mathrm{MHz}$ is the nuclear quadrupolar interaction constant, and $A=-2.16\ \mathrm{MHz}$ is the hyperfine coupling constant. $\omega_{e}$ ($\omega_{n}$) corresponds to the Zeeman frequency of the electron (nuclear) spin. The energy levels utilized in this experiment are represented by red bars in Fig.~\ref{fig2} (b) with $|...\rangle_e$ ($|...\rangle_n$) encoding the electron (nuclear) spin state. 
In this work, the electron spin was considered as the system and the nuclear spin served as an ancilla qubit to readout the ergotropy of the system state. 
In the construction of the model Hamiltonian $H_S$, $|0\rangle_e$ was mapped to the high-energy level $|e\rangle_S$ of $H_S$.
The magnetic field was set to $500\ \mathrm{G}$ and the NV center was polarized into state $|0\rangle_e|1\rangle_n$ via a green laser pulse\cite{Laser_Polarization} with the electron spin polarization being 0.97(1). 
The population of the nuclear spin state $|1\rangle_n$ is measured to be 0.99(1), which is near-unity in our experiment after the polarization (see details in the Supplementary Material).
As shown in Fig.~\ref{fig2} (b), microwave (MW) pulses represented by blue arrows and radio-frequency (RF) pulses represented by purple arrows were applied to manipulate the quantum states of the electron spin and nuclear spin, respectively.

The pulse sequence for the measurement of the ergotropy is shown in Fig.~\ref{fig2}(c). Practically, the ergotropy of a state is obtained by calculating the difference of the results of two energy measurements: the mean energy of the initial state and the mean energy of the state after an extraction operation is applied. Therefore, experimentally measuring the ergotropy requires implementing the pulse sequence twice. The whole sequence consists of four parts: polarization, state preparation, work extraction and energy measurement. The NV center was polarized into state $|0\rangle_e|1\rangle_n$ by a green laser pulse. The nuclear spin was prepared in an equal superposition state $(|0\rangle_n+|1\rangle_n)/\sqrt{2}$ to facilitate the measurement of the mean energy (see details in Supplementary Material)\cite{Measurement_Observable}. 
The preparation of the initial state and the following manipulation of the system state were realized by applying microwave pulses\cite{MW_Pulse1,MW_Pulse2}. 
The electron spin was first prepared in state $\rho_{S}$ by $U_P$.
The fidelity between the experimental initial state and the theoretical one is 1.00(1) (see details in Supplementary Material).
An extra free evolution time $t_D = 3T_{2}^{*}$ was required to dephase the coherence when preparing the dephased state $\delta_{\rho_S}$ with $T_{2}^{*}=56(3)\ \mathrm{\upmu}s$ being the dephasing time of the electron spin. The following pulses differ according to the target of the experiment trial: (i) to measure the mean energy of $\rho_{S}$ or $\delta_{\rho_S}$, the work extraction stage was only an identity evolution; (ii) to measure the mean energy of the states after work extraction, a proper operation $U_{ext}$ was applied in work extraction stage depending on which component of the ergotropy was to be measured (detailed pulses are in Supplementary Material). 
In our experiment, the operators are mainly affected by the spin-lattice relaxation, the interaction between the electron spin and the spin bath, and the fluctuation in the amplitude of controlling field. 
We estimated the fidelities of the operations in the experiment taking these effects into account. 
The fidelities between the experimental operations and the ideal operations are higher than 99\%, which shows that the operations in our experiment are very close to unitary (see detail discussion in the Supplementary Material).
Then a conditional unitary transformation $U_C=|1\rangle_n {}_{n}\langle 1|\otimes I + |0\rangle_n {}_{n}\langle 0|\otimes U_S$ was applied, where $U_S=e^{-iH_S \tau}$ with the evolution time $\tau=500\ \mathrm{ns}$. 
After the evolution time, the nuclear spin state is $\rho_n = (|0\rangle_n {}_{n}\langle 0|+|1\rangle_n {}_{n}\langle 1|+\mathrm{Tr}[{\rho_{f}U_S^{\dagger}}]|0\rangle_n {}_{n}\langle 1|+\mathrm{Tr}[{\rho_{f}U_S}]|1\rangle_n {}_{n}\langle 0|)/2$, where $\rho_{f}$ is the state of the electron spin before energy measurement. The information of the mean energy is encoded in the off-diagonal element of the nuclear spin state. Specifically, its imaginary part of the off-diagonal element is $\mathrm{Im}\mathrm{Tr}[{\rho_{f}U_S^{\dagger}}]\approx \mathrm{Tr}[{\rho_{f}H_S}]\tau=\langle H_S\rangle_{\rho_{f}}\tau$ with small evolution time approximation. 
This method to extract the mean energy can be applied to quantum systems of arbitrary dimension. 
Finally, another RF pulse was applied to measure the mean energy from the nuclear spin combined with the last laser pulse.    

\begin{figure}\centering
	\includegraphics[width=1\columnwidth]{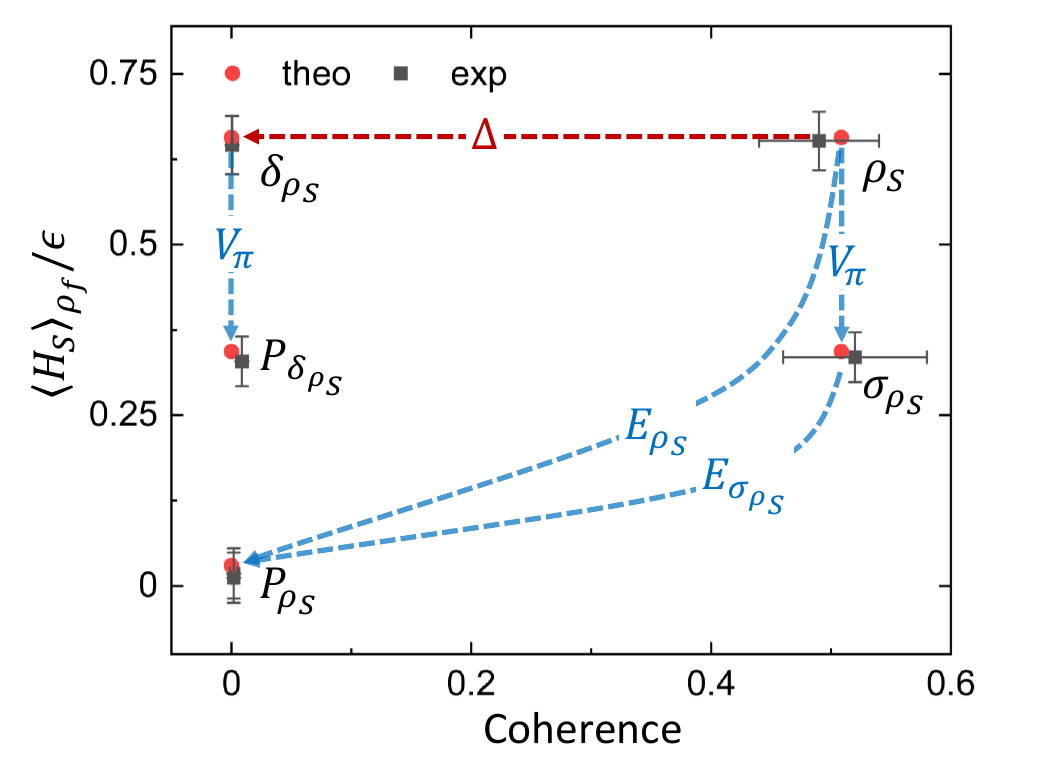}
	\caption{\label{fig3}
		Normalized mean energy and coherence of the system state before and after work extraction. Red dots (black squares) are the theoretical prediction (experimental result) of the mean energy and coherence. The influence of the imperfect polarization of the system spin is considered in the theoretical prediction. The green lines represent the the unitary work-extracting operations, the red line represent the dephasing process. Two experimental data points correspond to one theoretical point of $P_{\rho_{S}}$ because two trajectories lead to this point.
	}
\end{figure}

Fig.~\ref{fig3} displays the energy and coherence change with the initial state being $\rho_{S}$ or its totally dephased counterpart $\delta_{{\rho_S}}$. The influence of the imperfect electron spin polarization was considered in the theoretical prediction. The coherence was evaluated via state tomography. The ergotropy components can be calculated by subtracting the exhibited data points. The normalized incoherent ergotropy of $\rho_{S}$ is $\mathcal{E}_{i}({\rho_{S}})=\langle H_S\rangle_{{\rho_S}}-\langle H_S\rangle_{{\sigma_{\rho_S}}}=0.32(6)$, while the coherence remains unchanged in accordance with the definition of incoherent extraction. The totally dephased counterpart $\delta_{{\rho_S}}$ was transformed into its passive state by the same extraction operation $V_{\pi}$. The resulting incoherent optimal yield is $\mathcal{E}(\delta_{{\rho_S}})=\langle H_S\rangle_{{\delta_{\rho_S}}}-\langle H_S\rangle_{{P_{\delta_S}}}=0.32(6)$, which matches the value of the incoherent ergotropy of $\rho_{S}$. After the incoherent extraction of $\rho_{S}$, the resulting state $\sigma_{{\rho_S}}$ experienced one more extraction $E_{{\sigma_{\rho_S}}}$ to deliver the coherent ergotropy: $\mathcal{E}_{c}({\rho_{S}})=\langle H_S\rangle_{{\sigma_{\rho_S}}}-\langle H_S\rangle_{{P_{\rho_S}}}=0.32(5)$. Finally, the work was directly optimally extracted from $\rho_{S}$ to give the total ergotropy: $\mathcal{E}({\rho_{S}})=\langle H_S\rangle_{{\rho_S}}-\langle H_S\rangle_{{P_{\rho_S}}}=0.63(6)$. An alternative assessment that should give the same total ergotropy is summing the incoherent and coherent components of the ergotropy, which is $\mathcal{E}({\rho_{S}})=\mathcal{E}_{i}({\rho_{S}})+\mathcal{E}_{c}({\rho_{S}})=0.63(8)$. 
Our experimental results agree well with Eq.~\ref{Coh_Erg}.

\begin{figure}\centering
	\includegraphics[width=1\columnwidth]{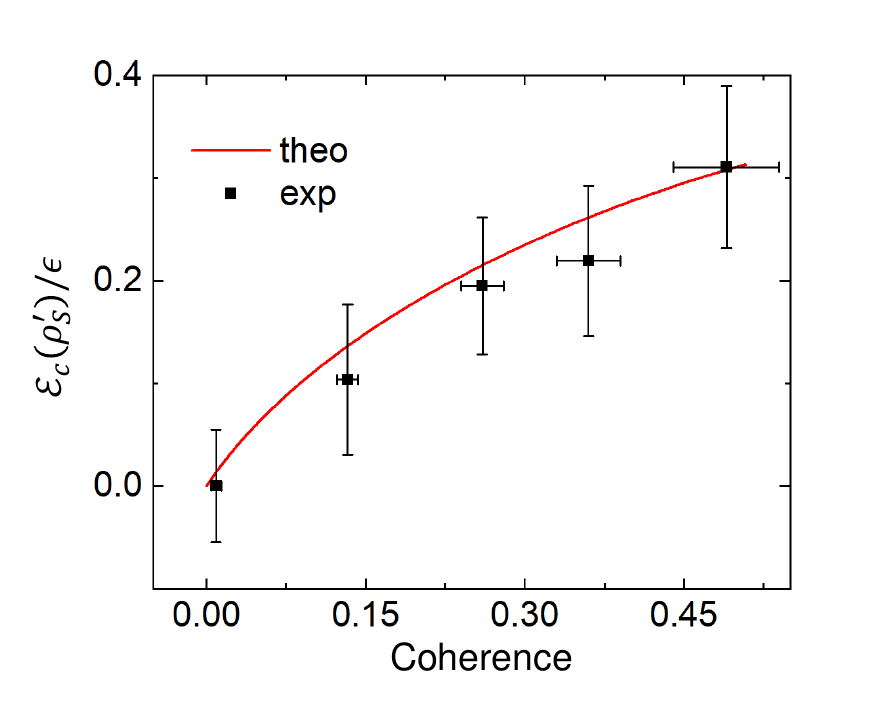}
	\caption{\label{fig4}
		The dependence of the normalized coherent ergotropy of the system state on the coherence. The black points are experimental date and the red solid line represents the theoretical prediction of the coherent ergotropy and coherence. Error bars show one standard deviation.
	}
\end{figure}

To further investigate the dependence of the coherent ergotropy on the coherence, we varied the coherence of the system state (see details in Supplementary Material) and measured its coherent ergotropy. 
We prepared state $\rho_{S}^{'}$, which has identical population distribution as $\delta_{\rho_{S}}$, but with different off-diagonal elements in energy basis. The increase of coherent ergotropy is given by $\mathcal{E}_c(\rho_{S}^{'})=(C(\rho_{S}^{'})-D(P_{{\rho_S^{'}}}||\rho_{\beta}))/\beta$. 
The coherence and coherent ergotropy of different $\rho_{S}^{'}$ were experimentally obtained. The theoretical prediction and experiment result are displayed in Fig.~\ref{fig4}.
For each state $\rho_{S}^{'}$, the optimal extraction operation was applied to evaluate the total ergotropy $\mathcal{E}(\rho_{S}^{'})$. The coherent ergotropy was obtained by taking the difference $\mathcal{E}_c(\rho_{S}^{'})=\mathcal{E}(\rho_{S}^{'})-\mathcal{E}_{i}(\rho_{S}^{'})$. The equality $\mathcal{E}_{i}(\rho_{S}^{'})=\mathcal{E}_{i}(\rho_{S})$ was utilized, since the incoherent ergotropy solely depends on the population distribution in energy basis. The experiment result aligns well with the theoretical prediction within one standard deviation. The positive contribution of the coherence to the ergotropy of a state is clearly confirmed.

In summary, we demonstrated the role of quantum coherence in the ergotropic work extraction in a single spin system. Both the incoherent and coherent ergotropy were experimentally measured, and the positive dependence of the coherent ergotropy on the coherence of the state is observed.
The interplay between quantum information theory and quantum thermodynamics is revealed by studying the work extraction process. Future studies could further investigate the relationship between the ergotropy of a quantum state and other distinctive quantum properties, such as entanglement\cite{Erg_Entanglement1,Erg_Entanglement2} and quantum discord\cite{Erg_QuantumSignatures}.
Additionally, the concept of the local and global ergotropy can also be studied by extending the system size\cite{Erg_Gap}.
The ergotropy has also been employed to study the thermodynamics of quantum systems subjected to environment, the work storing and extracting process can be studied by considering the effects of the surrounding environment\cite{Erg_Coh3,Erg_Open}.
Our work can potentially guide the enhancement of the capacity and efficiency of quantum devices operationally.

	\onecolumngrid
	\vspace{1.5cm}
	\begin{center}
		\textbf{\large Supplementary Material}
	\end{center}
	
	\setcounter{figure}{0}
	\setcounter{equation}{0}
	\setcounter{table}{0}
	\makeatletter
	\renewcommand{\thefigure}{S\arabic{figure}}
	\renewcommand{\theequation}{S\arabic{equation}}
	\renewcommand{\thetable}{S\arabic{table}}
	\renewcommand{\bibnumfmt}[1]{[RefS#1]}
	\renewcommand{\citenumfont}[1]{RefS#1}
	
	

\section{S1. Sample property}
\begin{figure}[ht]
	\centering
	\includegraphics[width=0.8\linewidth]{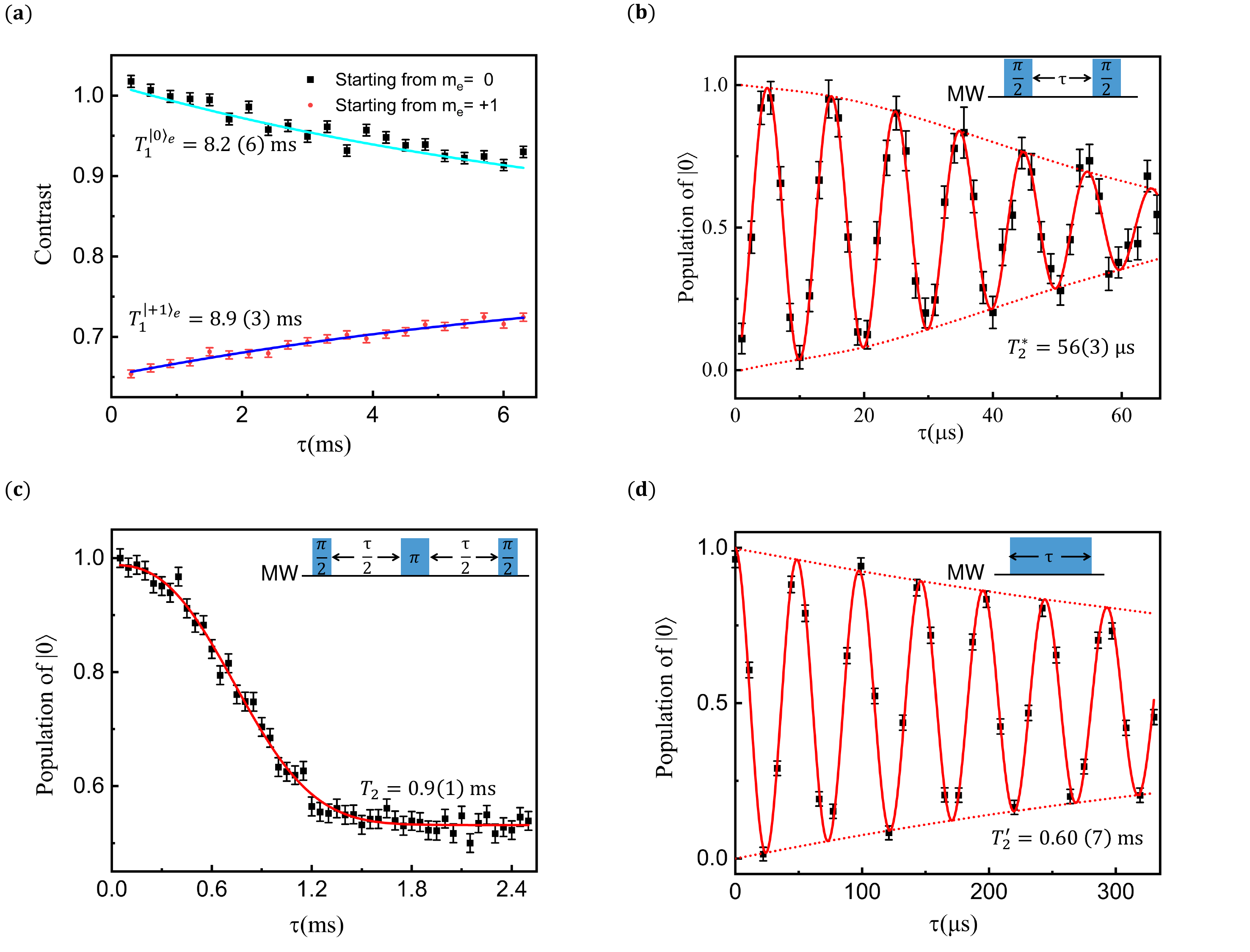}
	\caption{\label{RelaxationTimes} The relaxation times of the electron spin. (a) The spin-lattice relaxation times were measured with initial states being $|0\rangle_e$ or $|+1\rangle_e$, respectively. The solid lines are the fit to the experiment data (black squares and red dots). The results are $T_1^{|0\rangle_e}$=8.2(6) ms and $T_1^{|+1\rangle_e}$=8.9(3) ms. (b) Dephasing time of the electron spin measured by Ramsey	experiment (insert, pulse sequence). The solid red line is the fit to the experiment data (black square), and the red dashed line is the fit to the envelope curve. $T_2^*$ was measured to be 56(3) $\upmu\rm{s}$. (c) Result of $T_2$ for the electron spin by the spin echo experiment (insert, pulse sequence). The red solid line is the fit to the experiment data (black squares). $T_2$ was measured to be 0.9(1) ms. (d) Result of the nutation experiment (insert, pulse sequence) for the electron spin. The red solid line is the fit to the experiment data (black squares). The decay time of the nutation is $T_2'$= 0.60(7) ms. The error bars on the data points are the standard deviations from the mean value.}
\end{figure}

Our experiment was conducted on a nitrogen-vacancy (NV) center in the [100] face bulk diamond which was isotopically purified ([${}^{12}$C]=99.999\%). The electron spin qubit was chosen as the system qubit. 
The effects affecting the operations mainly come from the spin-lattice relaxation described by $T_1$, the interaction between the electron spin and the spin bath described by $T_2$ and $T_2^*$, and the fluctuation in the amplitude of controlling field described by $T_2'$. 
We have provided experiments to measure these relaxation times. 
The spin-lattice relaxation times with different initial states are both longer than 8 ms as shown in FIG.~\ref{RelaxationTimes}a. 
$T^*_2 = 56(3)\ \upmu\rm{s}$, $T_2$ = 0.9(1) ms and $T_2'$= 0.60(7) ms are obtained as shown in FIG.~\ref{RelaxationTimes}b,~\ref{RelaxationTimes}c and~\ref{RelaxationTimes}d, respectively. 
We estimated the fidelities of the operations in the experiment taking these effects into account. 
The fidelities between the experimental operations and the ideal operations are higher than 99\%, which shows that the operations in our experiment are very close to unitary.

\begin{figure}[ht]
	\centering
	\includegraphics[width=0.6\linewidth]{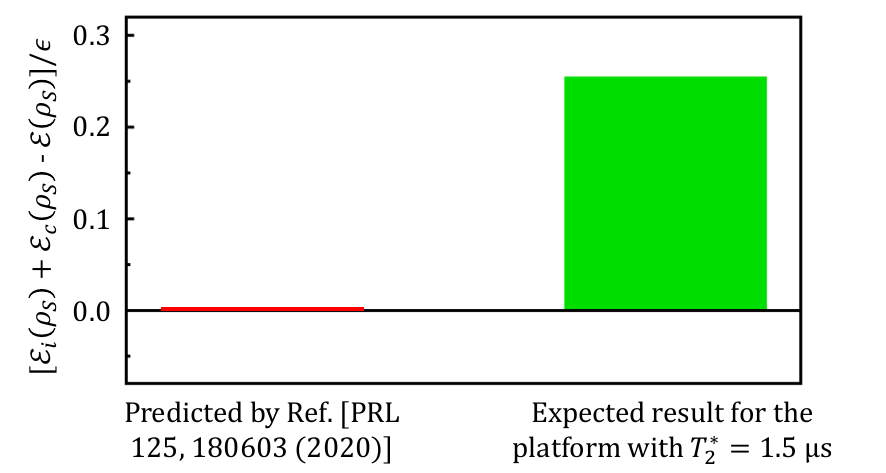}
	\caption{\label{ShortT2star} Verifying the conclusion in the reference [PRL 125, 180603 (2020)], $\mathcal{E}_{i}({\rho})+\mathcal{E}_{c}({\rho})-\mathcal{E}({\rho})=0$. The red bar is the theoretical prediction from the reference. The green bar is the expected result for the platform with $T^*_2 = 1.5\ \upmu\rm{s}$. }
\end{figure}

We also analysis the expected result based on NV center in diamond with natural abundance of $^{13}$C and we found that the relationship between ergotropy and coherence cannot be correctly reflected with such system.
The typical dephasing time of the electron spin is about 1.5 $\mathrm{\mu s}$ for the natural abundance diamond. 
As shown in FIG.~\ref{ShortT2star}, the crucial conclusion from the reference [PRL 125, 180603 (2020)], $\mathcal{E}_{i}({\rho})+\mathcal{E}_{c}({\rho})-\mathcal{E}({\rho})=0$, cannot be obtained using such sample.
To address this issue, we synthesized an isotopically purified diamond ([$^{13}$C]=0.001\%) with the electron spin dephasing time $T^*_2 = 56\ \upmu\rm{s}$. We ensured high-fidelity operations and successfully demonstrated the relationship between ergotropy and coherence.

\section{S2. Method of mean energy measurement}
\subsection{Theoretical description}
\begin{figure}
	\centering
	\includegraphics[width=0.5\linewidth]{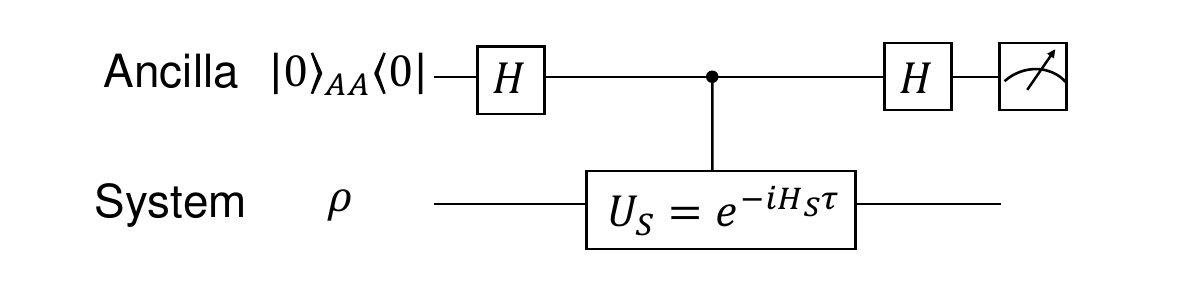}
	\caption{\label{Circuit} 
		The quantum circuit of the energy measurement.}
\end{figure}

By definition, determining the ergotropy of a quantum state requires measuring the mean energy of the state and that of the state after work extraction.
To measure the mean energy of a quantum state $\rho$ subjected to Hamiltonian $H_S$, we introduce an ancilla qubit\cite{Measurement_Observable}. The quantum circuit representation of the energy measurement is depicted in FIG.~\ref{Circuit}. Initially, the ancilla qubit and the system are prepared in state $\rho_{tot}=|0\rangle_{A}{}_{A}\langle 0|\otimes\rho$. Then the state of the ancilla qubit is transformed into an equal superposition state $(|0\rangle_{A}+|1\rangle_{A})/\sqrt{2}$. Afterwards, a conditional transformation $U_C=|0\rangle_{A}{}_{A}\langle0|\otimes I+|1\rangle_{A}{}_{A}\langle1|\otimes U_S$ is applied, where $U_S=e^{-iH_{S}\tau}$ is the evolution operator with evolution time $\tau$.
The state of the system and ancilla qubit becomes
\begin{equation}
	\rho_{tot}=U_{C}\rho U_{C}^{\dagger}=\frac{1}{2}
	\begin{pmatrix}
		\rho & \rho U_{S}^{\dagger}\\
		U_S\rho & U\rho U_{S}^{\dagger}\\
		
	\end{pmatrix}.
\end{equation}
The state of the ancilla qubit is
\begin{equation}
	\label{Ancilla state}
	\rho_{A}=\frac{1}{2}
	\begin{pmatrix}
		1 & \mathrm{Tr}[\mathit{\rho U_S^{\dagger}}]\\
		\mathrm{Tr}[\mathit{\rho U_S}] & 1\\		
	\end{pmatrix}.
\end{equation}
When the evolution time $\tau$ is small enough, the imaginary part of its off-diagonal element can be approximated as $\mathrm{Im}\mathrm{Tr}[\mathit{\rho U_S^{\dagger}}]\approx\mathrm{Tr}[\mathit{\rho H_S}]\tau=\langle H_S \rangle_{\rho}\tau$. 
Finally, the ancilla qubit is rotated so that its off-diagonal element can be obtianed by measuring the population distribution of the final state. 
In the above derivation process, there is no restriction on the quantum state $\rho$ and Hamiltonian $H_S$. Our method to extract the mean energy can be applied to quantum systems of arbitrary dimensions.

\subsection{Realization of the conditional transformation}
In our experiment, energy levels $|1\rangle_n|1\rangle_e$, $|0\rangle_n|1\rangle_e$, $|1\rangle_n|0\rangle_e$ and $|0\rangle_n|0\rangle_e$ were utilized. In the subspace spanned these energy levels, the Hamiltonian of the NV center can be simplified as
\begin{equation}
	H_{NV} = \pi[(Q+\omega_n+\frac{A}{2})\sigma_z\otimes I+(D+\omega_e+\frac{A}{2})I\otimes\sigma_z+\frac{A}{2}\sigma_z\otimes\sigma_z].
\end{equation}
In the construction of the model Hamiltonian $H_S$, $|0\rangle_e$ was mapped to the high-energy level $|e\rangle_S$ of $H_S$.
We choose the following interaction picture
\begin{equation}
	U_{rot} = e^{i(H_{NV}-|1\rangle_A{}_{A}\langle 1|\otimes H_S)t},
\end{equation}
where $H_S=diag(\epsilon,0)$ is the model Hamiltonian in the main text. The Hamiltonian of the NV center is transformed to 
\begin{equation}
	H_{rot} = U_{rot}H_{NV}U_{rot}^{\dagger}-iU_{rot}\frac{\partial U_{rot}^{\dagger}}{\partial t}= |1\rangle_A{}_{A}\langle 1|\otimes H_S.
\end{equation}
The conditional transformation $U_S$ is realized by subjecting the system to this Hamiltonian for time duration $\tau$.

\subsection{Experimental readout}
\begin{figure}[h]
	\centering
	\includegraphics[width=0.9\linewidth]{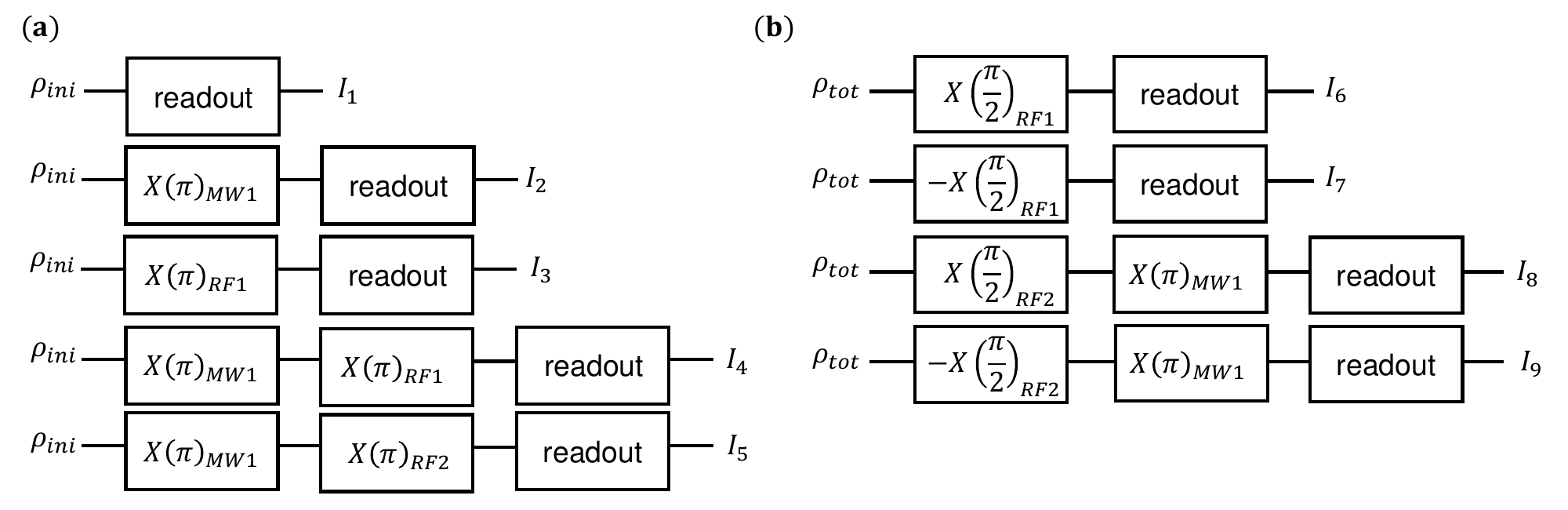}
	\caption{\label{Measurement Method}Sequence diagram for measurement of (a) the electron polarization and PL rates of the energy levels, (b) the off-diagonal element of the nuclear spin.}
\end{figure}

To experimentally obtain the mean energy of the system, the off-diagonal element of the nuclear spin was measured. The difference of the photoluminescence (PL) rates between the energy levels\cite{Brightness} we utilized can be used to readout the off-diagonal element of the nuclear spin. Thus, the pulse sequences shown in FIG.~\ref{Measurement Method} (a) were applied. The NV center was in state $\rho_{ini}=P_e|1\rangle_n|0\rangle_e+(1-P_e)|1\rangle_{n}|1\rangle_{e}$ after the application of the green laser pulse, where $P_e$ is the population of $|0\rangle_{e}$ after the optical polarization. We assumed that the nuclear spin polarization is 1. 
$X(\pi)_{MW1}$ ($X(\pi)_{MW2}$, $X(\pi)_{RF1}$, $X(\pi)_{RF2}$) denotes a selective $\pi$ pulse along $X$-axis between $|1\rangle_n|0\rangle_n$ and $|1\rangle_n|1\rangle_e$ ($|0\rangle_n|0\rangle_e$ and $|0\rangle_n|1\rangle_e$, $|1\rangle_n|0\rangle_e$ and $|0\rangle_n|0\rangle_e$, $|1\rangle_n|1\rangle_e$ and $|0\rangle_n|1\rangle_e$). We obtain the following equations via the pulses:
\begin{equation}
	\label{Cali_Eq}
	\begin{pmatrix}
		1-P_e & P_e & 0 & 0\\
		P_e   & 1-P_e & 0 & 0\\
		1-P_e & 0 & 0 & P_e\\
		P_e   & 0 & 0 & 1-P_e\\
		0     & P_e & 1-P_e & 0 
	\end{pmatrix}
	\begin{pmatrix}
		L_{|1\rangle_{n}|1\rangle_{e}}\\
		L_{|1\rangle_{n}|0\rangle_{e}}\\
		L_{|0\rangle_{n}|1\rangle_{e}}\\
		L_{|0\rangle_{n}|0\rangle_{e}}
	\end{pmatrix}
	=
	\begin{pmatrix}
		I_1\\
		I_2\\
		I_3\\
		I_4\\
		I_5
	\end{pmatrix}.
\end{equation}
$L_{|s\rangle}$ is the PL rate if the population of state $|s\rangle$ is 1, and $I_{k}$ is the PL rate obtained by the $k$-th sequence. By solving Eq.(\ref{Cali_Eq}), the PL rates can be obtained.

The state of the NV center after the conditional transformation takes the form
\begin{equation}
	\label{Total state}
	\rho_{tot}=
	\begin{pmatrix}
		\rho_{11} & \rho_{12} & \rho_{13} & \rho_{14}\\
		\rho_{21} & \rho_{22} & \rho_{23} & \rho_{24}\\
		\rho_{31} & \rho_{32} & \rho_{33} & \rho_{34}\\
		\rho_{41} & \rho_{42} & \rho_{43} & \rho_{44}
	\end{pmatrix}.
\end{equation}
Compare Eq.(\ref{Ancilla state}) and Eq.(\ref{Total state}), we have $\langle H_S\rangle_{\rho}=2\mathrm{Im}(\rho_{13}+\rho_{24})/\tau$. Thus, to measure the mean energy of the electron spin state, imaginary part of $\rho_{13}$ and $\rho_{24}$ are needed. By applying the pulse sequences in FIG.~\ref{Measurement Method} (b), we obtain the following equations
\begin{equation}
	\label{Offdiag}
	\begin{pmatrix}
		\rho_{11} & \frac{\rho_{22}+\rho_{44}}{2}+\mathrm{Im}\rho_{24} & \rho_{33} & \frac{\rho_{22}+\rho_{44}}{2}-\mathrm{Im}\rho_{24}\\
		\rho_{11} & \frac{\rho_{22}+\rho_{44}}{2}-\mathrm{Im}\rho_{24} & \rho_{33} & \frac{\rho_{22}+\rho_{44}}{2}+\mathrm{Im}\rho_{24}\\
		\rho_{22} & \frac{\rho_{11}+\rho_{33}}{2}+\mathrm{Im}\rho_{13} & \frac{\rho_{11}+\rho_{33}}{2}-\mathrm{Im}\rho_{13} & \rho_{44}\\
		\rho_{22}& \frac{\rho_{11}+\rho_{33}}{2}-\mathrm{Im}\rho_{13} & \frac{\rho_{11}+\rho_{33}}{2}+\mathrm{Im}\rho_{13} & \rho_{44}
	\end{pmatrix}
	\begin{pmatrix}
		L_{|1\rangle_{n}|1\rangle_{e}}\\
		L_{|1\rangle_{n}|0\rangle_{e}}\\
		L_{|0\rangle_{n}|1\rangle_{e}}\\
		L_{|0\rangle_{n}|0\rangle_{e}}
	\end{pmatrix}
	=
	\begin{pmatrix}
		I_6\\
		I_7\\
		I_8\\
		I_9
	\end{pmatrix}.
\end{equation}
By solving Eq.(\ref{Offdiag}), we obtain
\begin{equation}
	\begin{cases} 
		\mathrm{Im}\rho_{24}=\frac{I_6-I_7}{2(L_{|1\rangle_{n}|0\rangle_{e}}-L_{|0\rangle_{n}|0\rangle_{e}})}\\
		\mathrm{Im}\rho_{13}=\frac{I_8-I_9}{2(L_{|1\rangle_{n}|0\rangle_{e}}-L_{|0\rangle_{n}|1\rangle_{e}})} 
	\end{cases}.
\end{equation}
In the third and fourth sequences, an additional microwave (MW) pulse was inserted to enlarge the PL rate difference as $|L_{|1\rangle_{n}|0\rangle_{e}}-L_{|0\rangle_{n}|1\rangle_{e}}|\gg|L_{|1\rangle_{n}|1\rangle_{e}}-L_{|0\rangle_{n}|1\rangle_{e}}|$.

\section{S3. Experimental pulse sequences}
This section presents the pulse sequences used to experimentally investigate the coherent ergotropy of a quantum state. In subsection A, the detialed pulse sequences are displayed. In subsection B, examples of the waveforms of the MW pulses in subsection A are illustrated.
\subsection{Pulse sequences}
\begin{figure}[!h]
	\includegraphics[width=1\linewidth]{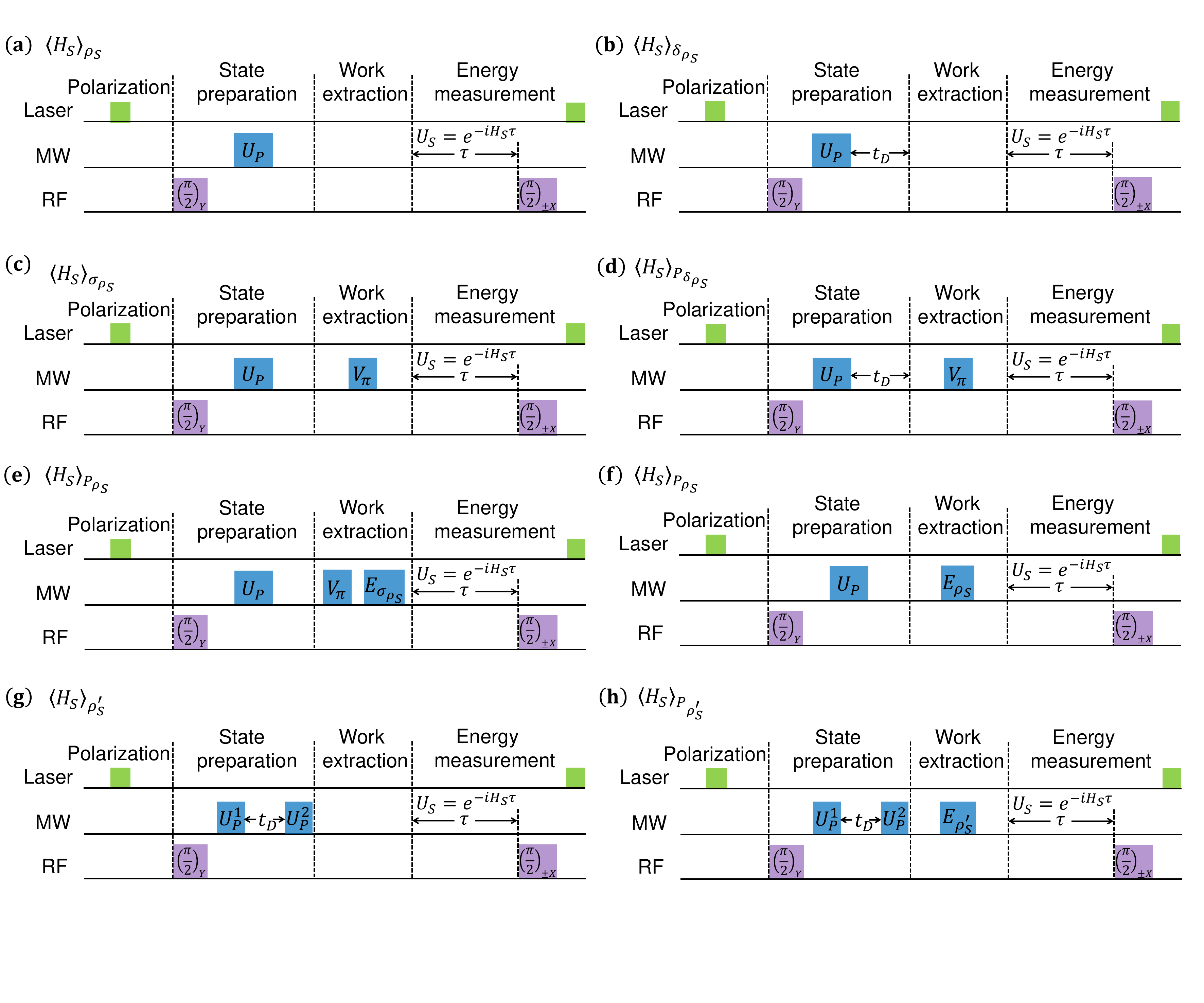}
	\caption{\label{Pulse Sequence}
		Experimental pulse sequences for the measurement of the ergotropy. The sequences consist of polarization, state preparation, work extraction and energy measurement. Each sequence correspond to the measurement of the mean energy of a quantum state: (a)$\rho_{S}$, (b)$\delta_{\rho_S}$, (c)$\sigma_{\rho_S}$, (d)$P_{\delta_{\rho_S}}$, (e)(f)$P_{\rho_S}$, (g)$\rho_{S}^{'}$, (h)$P_{\rho_S^{'}}$.
	}
\end{figure}

To experimentally investigate the coherent ergotropy, the electron spin of the NV center was taken as the system and the nuclear spin served as the ancilla qubit.
In the construction of the model Hamiltonian $H_S$, $|0\rangle_e$ was mapped to the high-energy level $|e\rangle_S$ of $H_S$.
The pulse sequences shown in FIG.~\ref{Pulse Sequence} were applied to manipulate the NV center. The electron spin and the nuclear spin were manipulated by MW and radio-frequency (RF) pulses, respectively.
Each pulse sequence consists of four steps: polarization, state preparation, work extraction and energy measurement. 
First, the NV center was polarized into state $|1\rangle_n|0\rangle_e$ by the green laser pulse. 
Second, the state of electron spin and nuclear spin were prepared. 
Third, a work-extracting operation was applied on the electron spin. 
Finally, the method introduced in Section II was implemented to measure the mean energy of the system state.

In our experiment, the total, incoherent, and coherent ergotropy were experimentally determined. This task involved measuring the mean energy of multiple states. Different pulses were applied in the state preparation and the work extraction stages to realize the mean energy measurement of different states as displayed in FIG.~\ref{Pulse Sequence}. 

\begin{figure}[!h]
	\centering
	\includegraphics[width=1\linewidth]{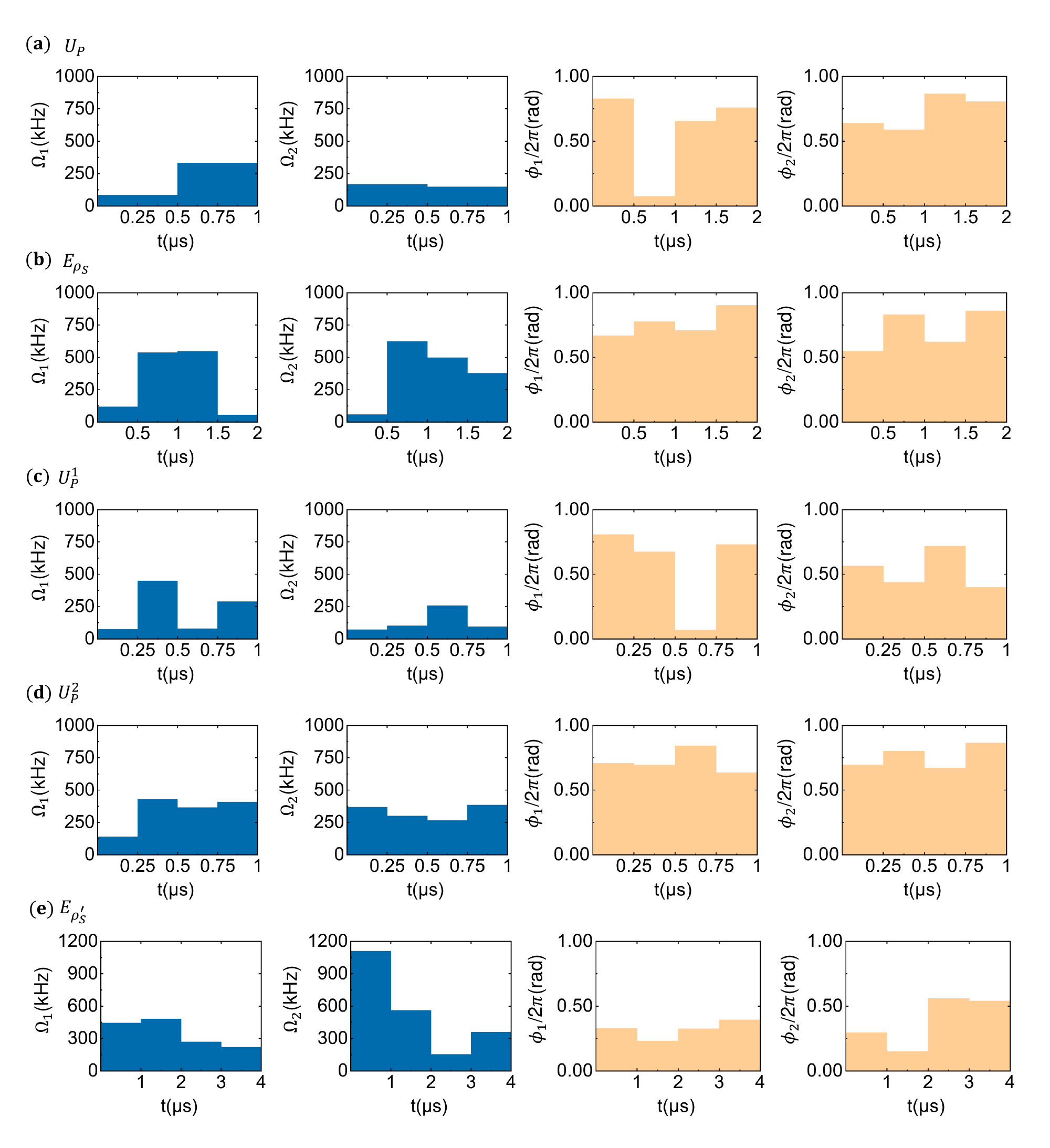}
	\caption{\label{Waveform}
		The amplitudes and phases of the optimized pulses: (a) $U_P$, (b) $E_{\rho_S}$, (c) $U_P^1$, (d) $U_P^2$ and (e) $E_{\rho_S^{'}}$. (c)-(e) correspond to the data point whose coherence is 0.13(1).}
\end{figure}

FIG.~\ref{Pulse Sequence} (a)-(f) display the pulse sequences used to obtain the incoherent, coherent and total ergotropy of $\rho_S$.
In the state preparation stage, $U_P=(\sqrt{2}|1\rangle_{e}{}_{e}\langle1|+|1\rangle_{e}{}_{e}\langle 0|-|0\rangle_{e}{}_{e}\langle 1|+\sqrt{2}|0\rangle_{e}{}_{e}\langle0|)/\sqrt{3}$ was applied to prepare the electron spin in state $\rho_{S}=|\psi_0\rangle\langle\psi_0|$, where $|\psi_0\rangle=(\sqrt{2}|0\rangle_e+|1\rangle_e)/\sqrt{3}$. 
In FIG.~\ref{Pulse Sequence} (b) and (d), an extra waiting time $t_D=3T_{2}^*$ was required in the state preparation stage to prepare the electron spin in $\delta_{\rho_S}=(2|0\rangle_e {}_e\langle 0|+|1\rangle_e {}_e\langle 1|)/3$. 
In FIG.~\ref{Pulse Sequence} (c)-(e), $V_{\pi}=|0\rangle_e{}_e\langle 1|-|1\rangle_e{}_e\langle 0|$ was applied to extract the incoherent ergotropy of the electron spin. 
In FIG.~\ref{Pulse Sequence} (e), $E_{\sigma_{\rho_S}}=(\sqrt{2}|1\rangle_{e}{}_{e}\langle1|-|1\rangle_{e}{}_{e}\langle 0|+|0\rangle_{e}{}_{e}\langle 1|+\sqrt{2}|0\rangle_{e}{}_{e}\langle0|)/\sqrt{3}$ followed $V_{\pi}$ to extract the coherent ergotropy of $\rho_S$.
In FIG.~\ref{Pulse Sequence} (f), $E_{\rho_S}=(|1\rangle_{e}{}_{e}\langle1|+\sqrt{2}|1\rangle_{e}{}_{e}\langle 0|-\sqrt{2}|0\rangle_{e}{}_{e}\langle 1|+|0\rangle_{e}{}_{e}\langle0|)/\sqrt{3}$ was applied to extract the total ergotropy of $\rho_S$. The sequences in both FIG.~\ref{Pulse Sequence} (e) and (f) were utilized to measure the mean energy of $P_{\rho_S}$, the passive state of $\rho_S$. FIG.~\ref{Pulse Sequence} (e) displays a two-step extractions in which the incoherent and coherent components of the ergotropy of $\rho_S$ were extracted consecutively, while FIG.~\ref{Pulse Sequence} (f) is the sequence that directly extracts the total ergotropy.

FIG.~\ref{Pulse Sequence} (g) and (h) display the pulses sequences to obtain the ergotropy of $\rho_S^{'}=\delta_{\rho_S}+(c/2)(|0\rangle_e{}_e\langle 1|+|1\rangle_e{}_e\langle 0|)$, where $c$ is a parameter to quantify the off-diagonal element of $\rho_{S}^{'}$. $\rho_S^{'}$ possesses identical population distribution as $\rho_S$ but different coherence. In the state preparation stage, $U_{P}^1=\cos{\frac{\alpha}{2}}(|0\rangle_e {}_e\langle 0|+|1\rangle_e {}_e\langle 1|)+\sin{\frac{\alpha}{2}}(|0\rangle_e{}_e\langle 1|-|1\rangle_e{}_e\langle 0|)$ was firstly applied to rotate the electron spin state, where $\alpha=\arctan{\sqrt{(8-9c^2)/(1+9c^2)}}$ is the rotation angle about Y-axis. Then a period of waiting time $t_D$ was required to dephase the coherence of electron spin state. Afterwards, $U_P^2=\cos{\frac{\theta}{2}}(|0\rangle_e {}_e\langle 0|+|1\rangle_e {}_e\langle 1|)+\sin{\frac{\theta}{2}}(|0\rangle_e{}_e\langle 1|-|1\rangle_e{}_e\langle 0|)$ was applied to further rotate the electron spin state so that the population distribution of the electron spin state $\rho_S^{'}$ is the same as $\rho_S$. The rotation angle of $U_P^2$ is $\theta=\arctan{3c}$ about Y-axis. The coherence of $\rho_S^{'}$ can be altered by varying the rotation angle of $U_P^1$. In FIG.~\ref{Pulse Sequence} (h), $E_{\rho_S^{'}}$ was applied to extract the total ergotropy of $\rho_S^{'}$. $E_{\rho_S^{'}}$ is the rotation about Y-axis whose rotation angle is $\pi-\theta$.

\subsection{Optimization of the pulses}

Implementing the energy measurement requires applying the same operations to the electron spin state in two subspaces where the nuclear spin state is $|1\rangle_n$ or $|0\rangle_n$. The main obstacle to achieving high-fidelity operations is the cross-talk between MW pulses of different frequencies. The optimal control method was employed to mitigate the cross-talk effect\cite{MW_Pulse1,MW_Pulse2}. Additionally, we also countered the dephasing and the fluctuation of the controlling field when we designed the shape of the pulses.


Two MW pulses were applied, the control Hamiltonian in the rotating frame is
\begin{equation}
	H_{C}(t)=\pi\Omega_1(t)I_n\otimes[\cos\phi_1(t)S_x+\sin\phi_1(t)S_y]+ \pi\Omega_2(t)I_n\otimes[\cos\phi_2(t)S_x+\sin\phi_2(t)S_y],
\end{equation}
where $\Omega_1(t)$ and $\Omega_2(t)$ are the amplitudes of the MW pulses, $\phi_1(t)$ and $\phi_2(t)$ are the phases.
The frequency of the MW pulse whose amplitude is $\Omega_1(t)$ ($\Omega_2(t)$) equals the energy difference between $|1\rangle_{e}$ and $|0\rangle_{e}$ with the nuclear spin state being $|1\rangle_n$ ($|0\rangle_n$). The amplitudes and phases were set as piecewise constants. The evolution time $T$, was equally split into $N=4$ segments. The evolution operator of the control Hamiltonian is $U(T)=e^{-i\int_{0}^{T}H_C(t)dt}$. The target operators were set as $I_n\otimes U_{targ}$, where $U_{targ}$ encompasses $U_P$, $V_{\pi}$, $E_{\sigma_{\rho_{S}}}$, $E_{\rho_S}$, $U_P^1$, $U_P^2$ and $E_{\rho_S^{'}}$. The fidelity between $U(T)$ and $U_{targ}$ is defined as $F=|\mathrm{Tr}[U^{\dagger}_{targ}U(T)]/\mathrm{Tr}[U^{\dagger}(T)U(T)]|^2$. The amplitudes and phases were optimized using gradient ascent pulse engineering algorithm. 

\section{S4. Results of initial state preparation}
\begin{figure}[!h]
	\centering
	\includegraphics[width=0.5\linewidth]{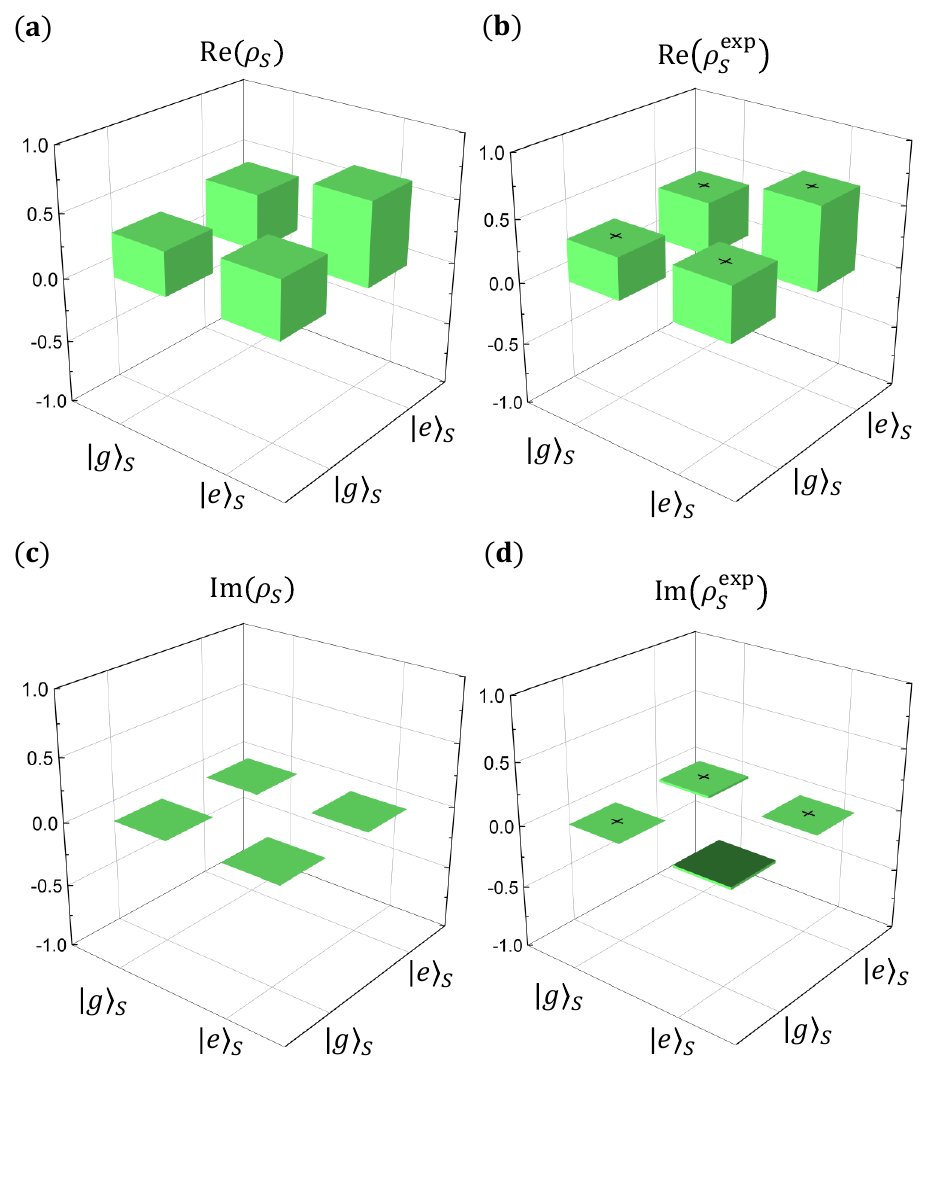}
	\caption{\label{InitialStateFidelity}
		(a) and (c) ((b) and (d)) are the real and imaginary parts of the ideal (experimental) density matrix of the initial state, respectively.}
\end{figure}

Through quantum state tomography, we experimentally obtained the density matrix of the initial state $\rho_S$ of the system as shown in FIG.~\ref{InitialStateFidelity}b and~\ref{InitialStateFidelity}d. The fidelity between the experimental initial state and the theoretical prediction is $F=\left[\mathrm{Tr}(\sqrt{\sqrt{\rho_S}\rho_S^{\mathrm{exp}}\sqrt{\rho_S}})\right]^2=1.00(1)$.

\begin{figure}[!h]
	\centering
	\includegraphics[width=0.5\linewidth]{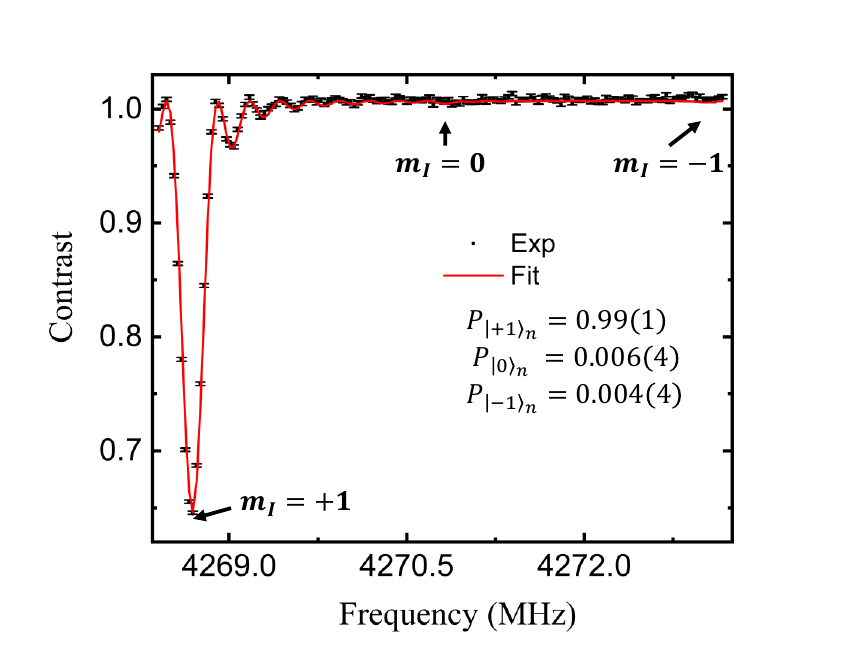}
	\caption{\label{ODMR}
		The experimentally obtained ODMR spectrum (black dots) and the fitted ODMR spectrum (red line).}
\end{figure}

The initial state of the nuclear spin as an ancilla qubit is characterized by the population of $|+1\rangle_{n}$.
The ODMR spectrum of the NV center was measured to obtain the population of $|+1\rangle_n$ via fitting the spectrum. The fitting result is $P_{|+1\rangle_{n}}=0.99(1)$ as displayed in FIG.~\ref{ODMR}, which shows that the population of $|+1\rangle_{n}$ is near-unity in our experiment. 

\end{document}